\documentclass[12pt,letterpaper]{article}
\usepackage{jheppub}

\title{Why Comparable? A Multiverse Explanation of the Dark Matter--Baryon Coincidence}

\author[a,b]{Raphael Bousso and Lawrence Hall}
\affiliation[a]{Center for Theoretical Physics and Department of Physics,\\
 University of California, Berkeley, CA 94720, U.S.A.}
\affiliation[b]{Lawrence Berkeley National Laboratory, Berkeley, CA 94720,
  U.S.A.}

\abstract{The densities of dark and baryonic matter are comparable: $\zeta\equiv \rho_D/\rho_B\sim O(1)$. This is surprising because they are controlled by different combinations of low-energy physics parameters. Here we consider the probability distribution over $\zeta$ in the landscape. We argue that the Why Comparable problem can be solved without detailed anthropic assumptions, and independently of the nature of dark matter. Overproduction of dark matter suppresses the probability like $(1+\zeta)^{-1}$, if the causal patch is used to regulate infinities.  This suppression can counteract a prior distribution favoring large $\zeta$, selecting $\zeta\sim O(1)$.  

This effect not only explains the Why Comparable coincidence but also renders otherwise implausible models of dark matter viable. For the special case of axion dark matter, Wilczek and independently Freivogel have already noted that a $(1+\zeta)^{-1}$ suppression prevents overproduction of a GUT-scale QCD axion. If the dark matter is the LSP, the effect can explain the moderate fine-tuning of the weak scale in simple supersymmetric models.}

\begin{document}
\maketitle

\section{Introduction}
\label{sec-intro}

Over the last few decades fundamental physics has been dominated by fine-tuning problems associated with the small scales of the cosmological constant, $\Lambda$, and the weak interactions, $v$.    Small scales can arise from different origins: naturally from symmetries, or by fine-tuning through environmental selection in a multiverse that scans the parameters that determine the scale.   Which mechanisms are at work for $\Lambda$ and $v$? 

A key difference is that no symmetries are known that could account for $\Lambda$, but supersymmetry or other new physics could stabilize the weak scale $v$.   The leading goal of the Large Hadron Collider (LHC) has been to search for a symmetry origin of $v$. Yet the discovery of a weakly coupled Higgs boson, without any signal of physics beyond the Standard Model so far, points to the absence of such a symmetry.  If this situation persists in coming runs at the LHC, it will become more plausible that not only $\Lambda$ but also $v$ is anthropically selected in a multiverse.  

In addition to these fine-tuning problems, fundamental physics must grapple with a variety of coincidence problems.  Two such problems are paramount in understanding the contents of the universe.  Matter and vacuum energy densities evolve differently, and yet they are comparable today: Why Now?  Moreover, while dark matter and baryon energy densities, $\rho_D$ and $\rho_B$, evolve according to the same power law, they have different origins and depend on different combinations of low-energy parameters. The two densities could easily differ by dozens of orders of magnitude, so it is remarkable that they are numerically so close~\cite{Planck13}:
\begin{equation}
\zeta\equiv \frac{\rho_D}{\rho_B} = \frac{n_D m_D}{n_B m_B} \approx 5.5~.
\label{eq:XoverB}
\end{equation}
Why Comparable?

New symmetries~\cite{Nussinov:1985xr, Gelmini:1986zz, Barr:1990ca, Barr:1991qn,Hooper:2004dc, Kaplan:2009ag} may explain why the number densities are comparable, $n_D/n_B \approx 1$. But the observed coincidence is in the mass density, so further work is required to link the dark matter and baryon masses, $m_D$ and $m_B$.  Meanwhile, cosmological observations and early LHC data increasingly favor multiverse solutions of the hierarchy problem, the cosmological constant problem, and the Why Now coincidence. Here we will argue that the multiverse can explain the Why Comparable coincidence, independently of the mass and nature of the dark matter particle. 

The probability distribution in the multiverse for observing a dark matter to baryon energy density $\zeta$ can be written as
\begin{equation}
dP = f(\zeta) \, \frac{d\zeta}{\zeta}\, \alpha(\zeta)\, M_b(\zeta)
\label{eq:genarg}
\end{equation}
where $f$ captures the distribution of the parameter $\zeta$ among the different metastable vacua in the landscape, $M_b(\zeta)$ is the total baryonic mass in regions with dark matter to baryon ratio $\zeta$, and $\alpha(\zeta)$ is the number of observations per unit baryonic mass in these regions. Because of the self-reproducing gravitational dynamics of metastable de~Sitter vacua, the total baryonic mass in the multiverse (and indeed, the total amount of any type of object or event whose probability is nonzero) diverges for any $\zeta$ and must be regulated. This is the measure problem of eternal inflation. It arises with the existence of at least one stable or metastable de~Sitter vacuum in the theory, such as (apparently) our own. Here we assume that the landscape has a very large number of vacua, at least enough to solve the cosmological constant problem~\cite{BP}.

The causal patch measure~\cite{Bou06,BouFre06a} is a theoretically well-motivated proposal that robustly solves the Why Now problem and predicts a value of $\Lambda>0$ in excellent agreement with observation~\cite{BouHar07}.\footnote{While entropic considerations were used in~\cite{BouHar07}, the resolution of the Why Now problem first identified there turned out to be independent of how observers are modeled. Only the causal patch measure plays an essential role.} These results follow from the geometry of the causal patch. They are insensitive to specific anthropic assumptions involving, say, the disruption of galaxy formation or other dynamical effects~\cite{BouHar10}.  The causal patch weights by the number of observations within a single event horizon.  If this region is mostly empty, as would be the case due to exponential dilution if the cosmological constant dominates before the era when observers live, then very little probability is assigned to the corresponding parameter range. 

Also working with the causal patch measure, Freivogel~\cite{Fre08} discovered the important result
\begin{equation}
M_b(\zeta) \propto \frac{1}{1+\zeta} ~.
\label{eq:gCD}
\end{equation}
He applied this result to axion dark matter, where the vacuum misalignment angle varies with a flat distribution, and found the observed dark matter abundance to be fairly typical.  Previously, again studying axion dark matter, Wilczek showed that by {\it assuming\/} a weighting of the form (\ref{eq:gCD}), the Why Comparable coincidence could be addressed~\cite{Wil04}.\footnote{The baryon fraction $f_b=(1+\zeta)^{-1}$ also appears in an early application~\cite{CliFre07} of the causal entropic principle~\cite{BouHar07} (and thus, in particular, of the causal patch measure). The example studied in Ref.~\cite{CliFre07} implicitly assumed a flat prior over $f_b$. This may be difficult to motivate, since $f_b\in [0,1]$ has finite range and does not depend simply on fundamental parameters.  In terms of $\zeta$, this prior takes the form $dN/d\zeta\propto (1+\zeta)^{-2}$. This is tantamount to {\em assuming\/} comparability of baryonic and dark densities: the prior already favors $\zeta\sim O(1)$. The measure factor and the catastrophic (entropic) boundary implicit in Ref.~\cite{CliFre07} sharpen this assumed preference.}

Here we argue that the causal patch measure allows a general solution of the Why Comparable problem, independently of the particle nature of dark matter.  The crucial point is that by Eq.~(\ref{eq:gCD}), the measure factor $M_b$ undergoes a sharp change in behavior in the vicinity of $\zeta = 1$.  The prior distribution $f$ is expected to be smooth in this region; the absence of a special scale dictates that $f \propto \zeta^n$.  This is the Why Comparable puzzle in the language of the landscape.  But the suppression of $M_b$ for $\zeta> O(1)$ leads to a maximum in $f M_b$ near $\zeta = 1$, if $0<n<1$. Satisfyingly, the ``anthropic factor'' $\alpha(\zeta)$ is not needed in this argument and can be set to a constant. If the number of observers per baryon drops for large or small values of $\zeta$, as has been argued~\cite{Lin88,TegAgu05,HelWal05,Hall:2011jd}, this will only improve an already satisfactory prediction.

The causal patch measure thus provides a unified and robust understanding of both the Why Now and Why Comparable coincidences: baryons, and therefore observations, must avoid being diluted by excess vacuum or dark matter energy density. Because the causal patch measure is defined geometrically and hence determined by the gravity of matter, it directly explains the coincidence of energy densities and not of number densities. Conventional anthropic assumptions are not needed.

\paragraph{Outline} In Sec.~\ref{sec-whycomp} we propose a multiverse explanation of the Why Comparable coincidence in a completely general setting. We combine a prior distribution over $\zeta$ with no special features at $\zeta\sim O(1)$ with the baryonic mass factor obtained from the causal patch measure. We obtain probability distributions with a broad peak around $\zeta\sim O(1)$. The observed value is typical in these distributions. We consider an interesting class of competing measure proposals and find that they do not lead to the same conclusion; so the causal patch is favored by our result.

In Sec.~\ref{sec-freeze}  we specialize to the case where dark matter is a long-lived particle with an abundance determined by thermal freeze-out in the early universe. We review the analysis linking its abundance to a mass scale near the weak scale.  Our solution of the Why Comparable problem generates this mass scale in the low-energy theory of observed vacua, independently of the weak scale and of the possible existence of new symmetries at that scale.  We thus relate the dark matter mass parametrically to the baryon abundance. It is only accidentally close to the weak scale, and generically slightly higher.

In Sec.~\ref{sec-little}, we specialize to an overlapping but different set of assumptions: that dark matter is the lightest supersymmetric particle (LSP).  We assume that the overall scale of supersymmetry breaking, $\tilde{m}$, is the only relevant scanning parameter, and study how the probability distribution for this scale is affected by the statistical preference for comparability,  $\zeta\sim O(1)$.  We consider both freeze-out relics and gravitinos. In both cases, we find that $\tilde m$ may not be far above the TeV scale even though it is unrelated to the weak scale. Solving the Why Comparable coincidence yields a fundamental reason for a little hierarchy.

\section{Solving the Why Comparable Coincidence}
\label{sec-whycomp}

In this section, we address the Why Comparable coincidence at a very general level. We do not assume the existence of catastrophic dynamical boundaries; for example, we do not assume that galaxy formation is adversely impacted when $\zeta$ becomes greater than some critical value. Parameters other than $\zeta$ will not enter the analysis; hence they can either be regarded as fixed to their observed values, or marginalized over. (Here we choose the latter option, in subsequent sections we will choose the former.) In Sec.~\ref{sec-cpm}, we show that the total baryonic mass $M_b$ depends on $\zeta$ in a very simple way determined by the geometry of the causal patch measure. In Sec.~\ref{sec-peak}, we combine this with the prior distribution over $\zeta$ among landscape vacua and show that under weak assumptions, the multiverse probability distribution over $\zeta$ has support mainly for $\zeta\sim O(1)$. We find in Sec.~\ref{sec-other} that this is not the case for other interesting measures.

\subsection{Suppression from the Causal Patch Measure}
\label{sec-cpm}

Consider a class of observers in the multiverse that exist at a fixed\footnote{Of course, the observed value $t_{\rm obs}\sim 10^{61}$ should not be very unlikely, if $t_{\rm obs}$ is allowed to scan. Perhaps the number of vacua in the landscape determine this value~\cite{BouFre10d}. If $t_{\rm obs}$ is correlated with $\zeta$, then marginalizing over $t_{\rm obs}$ will modify the prior probability distribution over $\zeta$. Our argument that $\zeta\sim O(1)$ can still be made, if appropriate versions of our (weak) assumptions on the prior, $f$, and on the number of observers per baryon, $\alpha$, are made directly on the overall probabilty distribution $f(\zeta) \alpha(\zeta) M(t_\Lambda, t_{\rm c})$, where $f$ is the marginalized prior distribution, $\alpha$ is the marginalized expected number of observers per baryon, and $M(t_\Lambda, t_{\rm c})$ is the weighted average matter mass after integrating out $t_{\rm obs}$.} time $t_{\rm obs}$ in a flat or open FRW universe, as would naturally be produced from the decay of a parent vacuum in the landscape, followed by a period of slow-roll inflation. One could set $t_{\rm obs}=13.7$ Gyr but this is not important for our argument. 

We will assume that the observers consist of baryonic matter.  We make no further assumptions about them, such as the need for galaxies, or carbon, etc.; we will assume instead that the number of observations per baryon at the time $t_{\rm obs}$ is fixed. That is, the total number of observations $N(\zeta)$ of a particular value of $\zeta$ is given by the total baryonic mass, $M_b(\zeta)$, inside the causal patch at the time $t_{\rm obs}$. 
\begin{equation}
N(\zeta)=\alpha M_b(\zeta)~.
\end{equation}
This will be sufficient to explain the Why Comparable coincidence, assuming only that the number of observers per baryon, $\alpha$, does not {\em increase\/} dramatically for large $\zeta$.

Previous multiverse analyses argued that observed values of $\zeta$ are limited by certain catastrophic dynamical boundaries, such as the failure to cool halos or to form stars~\cite{HelWal05,TegAgu05}.   It would be interesting if such a catastrophic boundary were close to the observed value of $\zeta \simeq 5.5$, especially if it provided an upper bound on $\zeta$.   However, there is no clear argument that such a nearby boundary exists.   Requiring early formation of halos with sufficient baryons to make at least one star yields a distant boundary, $\zeta \leq 10^5 - 10^6$,  and the requirement of disk fragmentation is highly uncertain, with a boundary plausibly in the range $\zeta \leq 10 - 10^4$, depending on the assumptions made.    These arguments are thus currently insufficient for explaining the Why Comparable coincidence. Here we show that they are not necessary either. The inclusion of presumed catastrophic boundaries in our analysis would sharpen the probability distribution without changing the main result, and/or allow us to further relax our (already weak) assumptions about the prior distribution over $\zeta$ in the landscape.

The class of vacua we consider is extremely broad: we will allow not only $\zeta$, but all low-energy parameters, to vary away from their observed values. There are some technical conditions we impose, which do not significantly impact the generality of our approach: 
\begin{itemize}
\item There exists a matter-dominated era.
\item The cosmological constant is not zero. (This is a technical assumption since the causal patch cutoff is not well-defined in vacua with $\Lambda=0$.\footnote{In fact, the causal patch and other leading cutoff proposals do not appear to be reliable in the causal neighborhood of big crunch singularities~\cite{BouLei09,BouFre10e}, so it may be appropriate to assume $\Lambda>0$. However, this is not necessary for our analysis.})
\item The time when observations are made, $t_{\rm obs}$, occurs no earlier than in the matter era.  It may lie in later eras (e.g., curvature or vacuum dominated), but we do not consider observers that exist during radiation domination.
\item We compute the probability distribution over the value of $\zeta$ at the time $t_{\rm obs}$. Dark matter that decays before $t_{\rm obs}$, or dark matter produced after $t_{\rm obs}$ is not constrained by our arguments.
\end{itemize} 

As a matter of principle, it is important to understand that the above conditions are not assumptions; our conclusions do not hinge on their universal validity. Perhaps there are observers that are not made from baryons, or which live in some other era. Here we compute conditional probabilities: what are typical observations made by observers in the class we have specified. Since this class includes ourselves, our approach could be falsified if we find that our observations are very atypical among such observers. Thus, our prediction that $\zeta\sim 1$ is a nontrivial success.

In fact, it would be legitimate to be far more restrictive, and to limit our attention to vacua that differ from ours only through the dark matter to baryon density ratio, $\zeta$. In the later sections of this paper, where we discuss concrete models of dark matter, we will indeed take this viewpoint. At the level of explaining the Why Comparable coincidence, however, considering a broader class of vacua does not complicate our task.  It allows us to claim a more general result that applies to all vacua with baryons and a matter-dominated era.  Any additional parameters can be scanned and integrated out. Our assumptions about the prior distribution for $\zeta$ refer to the marginalized distributions.

The key observation relevant to resolving the Why Comparable coincidence is extremely simple. The baryonic mass within the causal patch is given by
\begin{equation}
M_b=\frac{1}{1+\zeta} M(t_{\rm obs})~,
\label{eq-mbxil}
\end{equation}
where $M(t_{\rm obs})$ is the total matter mass in the causal patch at the time when observations are made. As reviewed in the Appendix, $M$ depends on $t_{\rm obs}$, on the cosmological constant, and on the time of curvature domination (if there is such an era). But in vacua satisfying the above rather weak conditions, {\em $M$ does not depend on $\zeta$}.   

This is the central point of our argument. It holds because the size of the causal patch is determined by computing the past light-cone of a point on the future boundary of the spacetime. Varying $\zeta$ could affect the time of equal matter and radiation density, but we have assumed that observations are made after this time. Because the patch is constructed from the future back, its size at $t_{\rm obs}$ is thus independent of $\zeta$.  In other words, $M$ depends only on parameters that are uncorrelated with $\zeta$. (For spatial curvature this could be considered a mild assumption.) Marginalizing over these parameters is thus trivial. $M$ does not depend on other particle physics parameters, such as the number of baryons per photon, which could introduce an implicit additional $\zeta$-dependence. Thus we find that the baryonic mass inside the patch depends on the dark matter to baryon ratio as
\begin{equation}
M_b\propto \frac{1}{1+\zeta}~.
\label{eq-mbxi}
\end{equation}
This equation is central to our solution of the Why Comparable problem.  We will now show that for a wide range of prior probability distributions with no special features near $\zeta\sim 1$, Eq.~(\ref{eq-mbxi}) leads to the prediction that $\zeta\sim O(1)$, independently of detailed anthropic assumptions.

\subsection{The Peak at $\zeta\sim 1$}
\label{sec-peak}

The prior distribution is the probability distribution over $\zeta$ in the ensemble of vacua produced by eternal inflation in the causal patch. Let
\begin{equation}
y=\log\zeta
\label{eq-mf1}
\end{equation}
and
\begin{equation}
f(y)=\frac{dN}{dy}~,
\label{eq-mf2}
\end{equation} 
where $dN$ is the number of vacua for which $\log\zeta$ lies in the interval $(y,y+dy)$. The derivative
\begin{equation}
F(y)=\frac{d\log f}{dy}
\label{eq-mf3}
\end{equation}
is called the prior multiverse force on $y$. 

We work with logarithmic variables because they make it simple to implement the requirement that no scale should be special in the prior distribution,\footnote{This is the weakest assumption one can start with. If a special scale was introduced into $f$ by hand, it would be easy to obtain $\zeta\sim 1$ but impossible to {\em explain} this coincidence.} by setting the prior multiverse force to a constant:
\begin{equation}
F=n~.
\end{equation}
Alternatively, one can think of this equation as a Taylor expansion of $\log f$ around some point of interest; in our case $y=0$ ($\zeta=1$). There may be corrections to the linear approximation, in violation of our earlier assumption that no scale is special in the prior distribution. A more minimal assumption is then that deviations from the linear approximation are not so drastic as to overcome the vast suppression of the probability distribution we obtain for the regime far from $y=0$. 

We will now show that for nearly any value of $n$ in the interval $(0,1)$, the overall probability distribution over $\zeta$ is peaked for values of order unity, explaining the coincidence that baryons and dark matter have comparable density. We will also show that the distribution is quite broad, so that the observed value, $\zeta=5.5$, is typical, i.e., in agreement with the theory. There is neither a surprisingly large nor a surprisingly small amount of dark matter in our universe.

\begin{figure}[tbp]\centering
\includegraphics[width=.9 \textwidth]{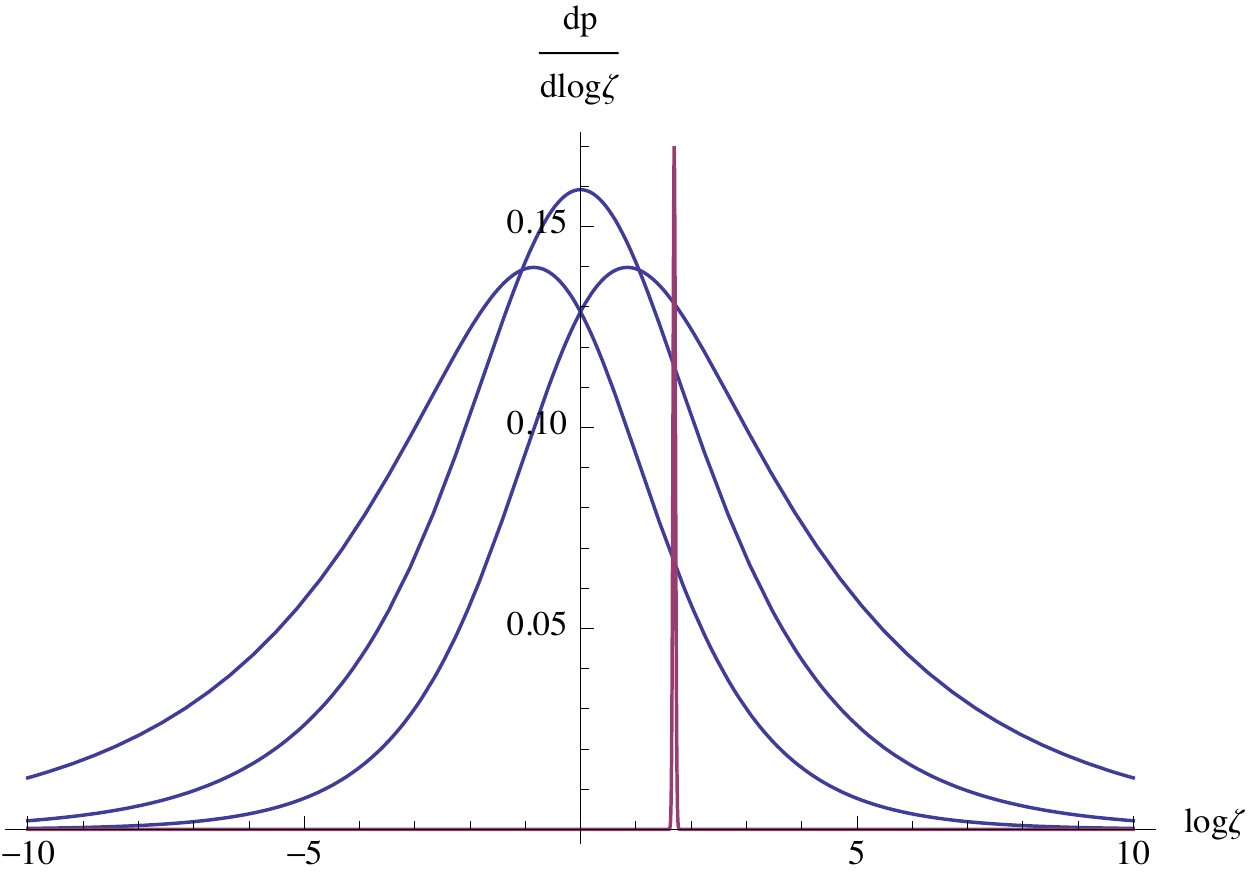}
\caption{Probability distribution over $y=\log\zeta$, for prior multiverse forces $n=.2$ (left), $n=.5$ (middle), and $n=.8$ (right). The left slope is due to the prior distribution favoring more dark matter; the right downslope is caused by the dilution of baryons in the measure factor. The observed value, $y\approx 1.7$, shown as a red spike, is quite typical in any of the three distributions.}
\label{fig-xiprobs}
\end{figure}

The probability distribution over $\zeta$ that is relevant to comparing theory with observation is not the prior, because values that are favored by the prior may not contain many observers. Rather, the probability distribution is proportional to the expected number of observations of the various possible values of $\zeta$ that are made inside the causal patch. This is given by the product of the prior---the probability that a particular value will actually appear as a vacuum region in the causal patch---with the number of observers inside the patch in such a vacuum, which is proportional to $M_b(\zeta)$. By Eq.~(\ref{eq-mbxi}) 
\begin{equation}
\frac{dp}{d\zeta}\propto\frac{\zeta^{n-1}}{1+\zeta}~,
\label{eq-zetadist}
\end{equation}
or in terms of the logarithmic variable:
\begin{equation}
\frac{dp}{dy}\propto\frac{e^{ny}}{1+e^y}~.
\end{equation}
This distribution is shown for three values of $n\in (0,1)$ in Fig.~\ref{fig-xiprobs}.

The distribution has a peak near $y=0$ ($\zeta\sim 1$). For large negative values of $y$, the multiverse force is positive, $F=n$, favoring more dark matter. But for large positive values it is negative, $F=n-1$, favoring baryons.   This is easily understood: the prior distribution is rising for $y<0$, whereas the measure factor $(1+\zeta)^{-1}$ is nearly constant and so does not affect the prior much in this regime. But for positive values of $y$ ($\zeta\gg 1$), the measure factor becomes important, and overwhelms the prior. Because the density drops off rapidly away from $y=0$, the values of $\zeta$ that are most likely to be observed are of order unity. This explains the Why Comparable coincidence.

\begin{figure}[tbp]\centering
\includegraphics[width=.9 \textwidth]{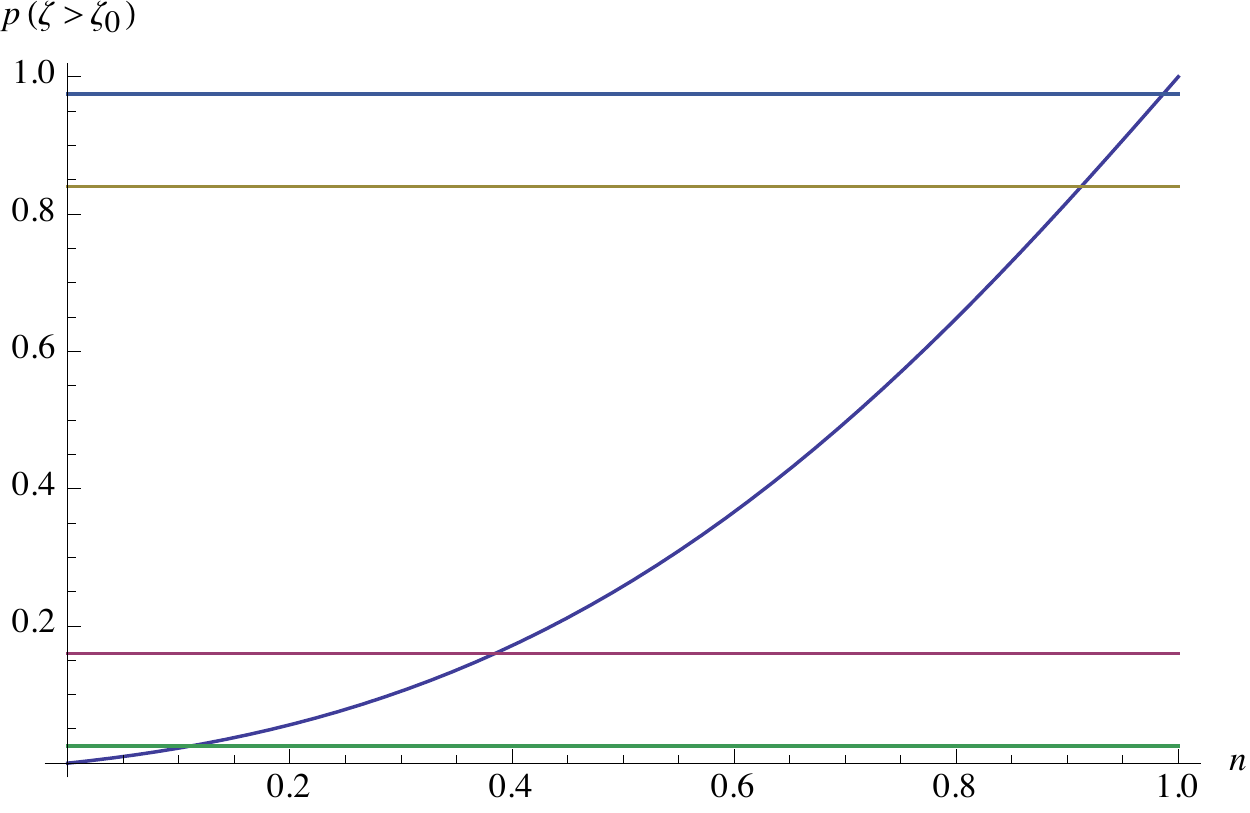}
\caption{As a function of the prior multiverse force $n$, the probability is shown for observers to find themselves in vacua with more dark matter than the observed value, $\zeta>5.5$. For our vacuum to lie within the central $1\sigma$ of the probability distribution over $\zeta$ (horizontal lines), $n$ must lie between about $.38$ and $.91$. For nearly all values of the multiverse force in the range shown, our observation of $\zeta=5.5$ is at least within the central $2\sigma$ of the predicted distribution. For values of $n$ outside the interval $(0,1)$, the probability distribution would peak at $\zeta\to 0$ or $\zeta\to \infty$, and our observation would be very unlikely unless explicit anthropic assumptions are introduced into the model.}
\label{fig-nrange}
\end{figure}

The above argument applies to any value of $n\in (0,1)$, but for values very close to the boundary, the smallness of $n$ or of $(1-n)$ introduces a new small parameter into the problem, and the peak need no longer lie at $\zeta\sim 1$.  One finds, however, that the observed value of $\zeta$ is typical (i.e., within the central $2\sigma$ of the distribution) for nearly all values of $n\in (0,1)$; see Fig.~\ref{fig-nrange}.

A prior landscape force $n<0$ would imply that most vacua have very little dark matter; a prior force $n>1$ would mean that such a large fraction of vacua have large $\zeta$ that the measure factor cannot overcome this pressure. In either case, the observed value would be very unlikely unless one assumes explicit catastrophic boundaries that cut off the probability distribution, such as a failure to form galaxies~\cite{TegAgu05}.   Moreover, one cannot understand the Why Comparable coincidence in this way.

For $n\in (0,1)$, remarkably, no such assumptions are needed, $\zeta=5.5$ is typical, and the Why Comparable coincidence is explained. Moreover, it is not difficult to obtain $n\in (0,1)$ from plausible assumptions about the landscape. A particularly compelling example is due to Freivogel~\cite{Fre08}: a natural (GUT or Planck scale) QCD axion~\cite{PecQui77}, which contributes a dark matter abundance of
\begin{equation} 
\zeta\sim \left(\frac{f_a}{10^{12}~\mbox{GeV}}\right)^{7/6} \langle\theta_i\rangle^2~.
\label{eq-axion}
\end{equation}
The main assumption in this case is a low scale of slow-roll inflation in the relevant class of vacua. The energy scale of inflation must be lower than the Peccei-Quinn symmetry breaking scale, $f_a$. (It must also be low enough to evade constraints on isocurvature perturbations.) Then the axion misalignment angle $\theta_i$ at the end of inflation is random and constant over the scale of the horizon at $t_{\rm obs}$.  Thus, the prior distribution over $\theta_i$ is flat, and by Eq.~(\ref{eq-axion}), the prior over $\zeta$ is
\begin{equation}
\frac{dN}{d\zeta}\propto \frac{d\theta}{d\zeta}\propto \frac{1}{\sqrt{\zeta}}~.
\end{equation}
By Eqs.~(\ref{eq-mf1})--(\ref{eq-mf3}), this corresponds to a prior multiverse force $n=\frac{1}{2}\in (0,1)$. 

This result is important for two reasons. It renders a high-scale QCD axion viable both as dark matter and as a solution to the strong CP problem, without the need for the controversial assumption that too much dark matter would prevent efficient structure formation~\cite{TegAgu05,HelWal05}. Just as importantly (though this aspect was not emphasized in Ref.~\cite{Fre08}), it explains the Why Comparable coincidence, if the dark matter is an axion. 

From the viewpoint we have developed above, these two features will be shared by any landscape model with prior pressure $n\in (0,1)$ towards large dark matter abundance. Such models would overproduce dark matter, were it not for the measure factor suppressing large $\zeta$. And with the measure factor taken into account, overproduction is averted and the Why Comparable coincidence is explained. In the following sections, we will illustrate this point by considering the dark matter as a freeze-out relic, and the dark matter as the LSP. 

We note briefly that our solution of the Why Comparable problem generalizes if dark matter has multiple components.  Each component has a density parameter $\zeta_i = \rho_{Di}/\rho_B$ distributed according to some a prior distribution $dN = f(\zeta_i) d \log \zeta_i$ and corresponding force $F_i =  d \log f / d \log \zeta_i = n_i$. The force for $\zeta = \Sigma \, \zeta_i$ is then given by $F=n\equiv \Sigma' n_i$, where the prime indicates that only terms with positive $n_i$ are included in the sum. Our solution to the Why Comparable problem requires $n<1$, so that it may be unlikely that there are a large number of dark matter components having $n_i>0$. However, there could certainly be a few such components and these typically have roughly comparable $\rho_{Di}$.  For example, if dark matter has both an axion and a freeze-out component then $n=0.5 + n_{FO}$, where $0<n_{FO}<0.5$ is the force on the freeze-out relic density, which could result from a force on the dark matter mass between 0 and 1.  

\subsection{Other Measures} 
\label{sec-other}

Before moving on, we note that the above result is specific to the causal patch measure. Other popular measures do not lead to Eq.~(\ref{eq-mbxi}). As an example, consider the fat geodesic cutoff~\cite{BouFre08b}, which is closely related to an interesting class of local~\cite{BouMai12} and global~\cite{DesGut08a,Bou12b} measures. We may analyze this measure in its local formulation. Consider an ensemble of geodesics orthogonal to a fiducial initial volume in the early universe. Each geodesic is thickened by a fixed infinitesimal physical volume $\delta V$. (For equivalence to the scale factor cutoff, this volume must be representative of the attractor regime of eternal inflation, but the present analysis holds for more general initial conditions.) The fat geodesic measure contains baryonic mass $M_b^{\rm FG}=\rho_b\delta V$, where $\rho_b$ is the density of baryons at the time of observation, $t_{\rm obs}$. 

The causal patch is the largest causally connected region in the universe and effectively averages $\rho_b$ over a large volume; it depends on the total mass within the event horizon, not how it is distributed. But the fat geodesic is sensitive to the density of matter at its own location.\footnote{For the same reason, this class of measures does not solve the Why Now problem~\cite{BouFre08b}, for positive values of $\Lambda$.} Timelike geodesics behave like collisionless dark matter and thus trace the dark matter. They will end up in structures with virial density $\rho_{\rm vir} =Q^3 T_{\rm eq}^4$.  Here $Q\sim 10^{-5}$ is the primordial density contrast, and $T_{\rm eq}\sim \xi_b+\xi_D=\xi_b(1+\zeta)$ is the temperature at matter-radiation equality; $\xi_b$ and $\xi_D$ are the baryonic and dark matter mass per photon. Hence the baryonic mass inside the fat geodesic at $t_{\rm obs}$ is
\begin{equation}
M_b^{\rm FG}\propto \frac{\rho_{\rm vir} }{1+\zeta}\propto Q^3\xi_b^4 (1+\zeta)^3 ~.
\label{eq-mfg}
\end{equation}
(Marginalizing over the cosmological constant~\cite{BouFre08b} would contribute a further factor of $\rho_{vir} \sim Q^3 \xi_b^4 (1 + \zeta)^4$ to the above formula, providing an even stronger force to large $\zeta$.) Since $Q$ and $\xi_b$ might depend on $\zeta$, further analysis requires more assumptions than for the causal patch measure. In fact, we have already assumed implicitly that structure forms, which was not necessary for the causal patch. 

It will suffice to show in a particularly natural setting that the fat geodesic measure cannot predict $\zeta\sim O(1)$. Suppose that the physics of baryogenesis and inflation is held fixed, so that $Q$ and $\xi_b$ are independent of $\zeta$. This is natural when studying the abundances of axion dark matter as in Ref.~\cite{Fre08}, or of the lightest supersymmetric particles, as in Sec.~\ref{sec-little}. By Eq.~(\ref{eq-mfg}), large dark matter abundance is strongly favored, because it increases the baryonic density near the fat geodesic. This cannot be compensated by assuming that vacua with large dark matter abundance are sufficiently rare ($n$ sufficiently small in the notation of the previous subsection): if the prior were strong enough to overcome a force towards large $\zeta$ which is strongest for $\zeta\gg 1$, then it would push right past $\zeta\sim O(1)$. Unless more specific anthropic assumptions are added by invoking catastrophic dynamical boundaries, the fat geodesic predicts that $\zeta$ should either be much less than or much greater than unity, for any prior without special features at $\zeta\sim O(1)$.

\section{Origin of the Mass Scale of Freeze-Out Dark Matter }
\label{sec-freeze}

A cosmological relic from thermal freeze-out is frequently called  a Weakly Interacting Massive Particle, or WIMP, because the mass that yields the observed abundance appears to be close to the weak scale, and because the hierarchy problem might be solved by assuming a new symmetry at this scale. In this section, we show that the mass of a freeze-out relic is connected, via our solution to the Why Comparable problem, to the baryon density of the universe. Its proximity to the weak scale is accidental and, by construction, unrelated to the hierarchy problem. For natural choices of couplings, values in the multi-TeV domain, somewhat above the weak scale, appear to be favored.

\subsection{The Mass Scale of a Thermal Relic}
\label{sec-relic}

We begin with a brief review of thermal freeze-out, following Ref.~\cite{Fen10}. We assume a particle with lifetime exceeding $t_{\rm obs}$ and mass  $m_X \leq T_r$, the reheat temperature after inflation, having sufficient interactions that it is brought into thermal equilibrium by temperature $T \sim m_X$.  As the temperature drops below $m_X$, as long as annihilation is efficient, the number density becomes Boltzmann suppressed:
\begin{equation}
n \sim (m_X T)^{3/2} e^{-m_X/T}~.
\label{eq-numd}
\end{equation}
The particles become too dilute to annihilate when the mean free time to annihilation becomes longer than the Hubble time (``freeze-out''):
\begin{equation}
n \langle \sigma_A v \rangle \sim H~,
\end{equation}
where $\sigma_A$ is the annihilation cross-section and $v$ is the velocity of dark matter particles. The thermally averaged annihilation cross-section is of the form
\begin{equation}
\langle\sigma_Av\rangle = \frac{c}{m_X^2}~,
\end{equation}
where $c$ involves a product of coupling constants, and may depend on mass ratios.   With $H\sim T^2/M_{\rm P}$, the freeze-out temperature satisfies
\begin{equation}
\left(\frac{T_f}{m_X}\right)^{1/2} e^{m_X/T_f} \sim c \frac{M_{\rm P}}{m_X}~.
\label{eq-tf}
\end{equation}
Equivalently, with $x_f\equiv m_X/T_f$,
\begin{equation}
x_f - \frac{1}{2}\log x_f\sim \log\left( \frac{cM_{\rm P}}{m_X}\right)~.
\label{eq-xf}
\end{equation}
Note that $x_f$ depends very weakly on the dark matter mass. For $m_X$ within one or two orders of magnitude of the TeV scale, justified below, and typical values of $c$, one finds $x_f\approx 20$. 

It is useful to express the dark matter abundance in terms of an approximately conserved quantity, such as the dark matter mass per photon:
\begin{equation}
\xi_X\sim \frac{m_X n}{T^3}=\frac{m_X n_f}{T_f^3}~.
\end{equation}
With Eqs.~(\ref{eq-numd}) and (\ref{eq-tf}) holding at freeze-out, one finds the well-known result for the relic abundance from thermal freeze-out 
\begin{equation} 
\xi_X \sim\frac{x_f }{\langle\sigma_Av\rangle M_{\rm P}}    \sim\frac{x_f }{cM_{\rm P}} m_X^2~.
\label{eq-zetaFO}
\end{equation}

\subsection{The Why Comparable Coincidence Sets the Scale}
\label{sec-setting}

For our purposes, it is crucial to compare this to the baryon mass per photon, which is given at late times by
\begin{equation}
\xi_b \sim m_p \eta~,
\end{equation}
where $\eta=n_b/n_\gamma\approx 6\times 10^{-10}$ is the baryon asymmetry and $m_p$ is the proton mass. The dark matter to baryon ratio is thus given by
\begin{equation}
\zeta=\frac{\xi_X}{\xi_b}\sim \frac{x_f}{c\,\eta} 
\frac{m_X^2 }{M_{\rm P }m_p}~.
\label{eq-zetaFO2}
\end{equation}
The ratio grows quadratically with $m_X$, apart from the weak logarithmic dependence determined by the transcendental Eq.~(\ref{eq-xf}).  

We now assume that $m_X$ varies in the landscape, with a probability distribution that has no preferred scale and is described by a probability force $n/2$, with $n\in (0,1)$ favoring large values of $m_X$.  Then $\zeta$ has a probability force $n$ and, taking other parameters fixed, by the analysis of the previous section the distribution for $\zeta$ is peaked near unity. As shown in Fig.~\ref{fig-xiprobs}, the observed dark matter abundance, $\zeta=5.5$, is quite typical. 

This result leads to a statistical prediction for $m_X^2$:  
\begin{equation}
m_X^2 \sim x_f^{-1} M_{\rm P} m_p \, c\, \eta .
\label{eq-FOscaling}
\end{equation}
With order-one factors restored and the dependence on $\zeta\sim O(1)$ made explicit, this prediction becomes
\begin{equation}
m_X = 0.5 \left( \frac{\alpha_{\rm eff}}{0.01} \right)  \sqrt{\zeta} \; \mbox{TeV},~~\zeta\sim O(1),
\label{eq-mXFO}
\end{equation}
where we have defined an effective coupling strength by $c= 4 \pi \alpha_{\rm eff}^2$, and normalized to the value that results for freeze-out of a vector electroweak doublet fermion annihilating to gauge bosons, $\alpha_{\rm eff} = 0.01$.  

Of course, inserting the observed value, $\zeta\approx 5.5$, in (\ref{eq-mXFO}) gives the usual thermal freeze-out result, $m_X\approx 1$ TeV.  The key point is that we have explained {\em why\/} $\zeta\sim O(1)$, the Why Comparable coincidence. We conclude that the mass scale of the freeze-out relic, 
\begin{equation}
m_X \sim  \alpha_{\rm eff} \,  \sqrt{\eta \, M_{\rm P} m_p }
\end{equation}
is set by the statistical preference for $\zeta\sim O(1)$.  It is parametrically unrelated (though accidentally close) to the weak scale. In Sec.~\ref{sec-LSP}, we will apply this result to explain why the SUSY breaking scale may be close to, but somewhat above, the weak scale. 

\section{The Origin of a Little Supersymmetric Hierarchy}
\label{sec-little}

In this section, we explore our explanation of the Why Comparable coincidence in theories with supersymmetric dark matter. We will begin, in Sec.~\ref{sec-hierarchy}, by considering the implications of the hierarchy problem for the expected scale of SUSY breaking in the landscape. We will argue that without taking dark matter into account, there are only two simple possibilities: either, the weak scale is natural, in which case SUSY should have been already discovered. Or SUSY should be broken far above the weak scale and remain out of reach. If dark matter is the LSP, however, we find that the baryon dilution factor $(1+\zeta)^{-1}$ can make it statistically likely for SUSY to be broken near the weak scale without rendering it natural. We consider two classes of models in detail: thermal freeze-out of a standard model superpartner, in Sec.~\ref{sec-LSP}, and gravitino LSPs produced by various mechanisms in Sec.~\ref{sec-gravitino}. In all cases, one expects a small hierarchy between the weak scale and the scale of observable superpartners.

\subsection{Naturalness and the Prior for $\tilde m$ }
\label{sec-hierarchy}

In a wide class of supersymmetric theories, the scale of weak interactions, $v$, is related to the overall mass scale of Standard Model superpartners, $\tilde{m}$, by
\begin{equation}
v^2 = (C_1 + C_2 + ...) \, \tilde{m}^2.
\label{eq-vtm}
\end{equation}
The parameters $C_i$ depend on details of the model, such as coupling constants and mass ratios. But generically they should be of order unity. Absent fine-tuning of $\sum C_i$, supersymmetry can stabilize the weak scale against radiative corrections only if the superpartners have mass near the observed weak scale.

In this section, we assume that the mass scale $\tilde{m}$ scans in the landscape, with a prior distribution
\begin{equation}
\frac{dp}{d \log \tilde{m}} \; \propto \; \tilde{m}^q
\end{equation}
with $q>0$.  This preference for large values of $\tilde{m}$ in the prior distribution need not lead to a prediction of large {\em observed} values. With all other model parameters held fixed, increasing $\tilde m$ drives up the weak scale, by Eq.~(\ref{eq-vtm}). All compound nuclei are unstable if $v$ exceeds the value we observe by more than 60\% \cite{AgrBar97}. We assume that this suppresses the abundance of observers dramatically, so that $v\sim 1.6 v_o$ can be regarded as a catastrophic boundary.

However, one expects that parameters of the supersymmetric model do vary in the landscape, in such a way that one or more of the $C_i$ scan. Then $\tilde{m}\gg v$ can occur, provided a cancellation allows $(C_1 + C_2 + ...) =v^2/\tilde m^2\ll 1$.  The prior probability for such a cancellation (i.e., the probability that it occurs in a randomly chosen vacuum) can be estimated by noting that $\sum C_i=0$ should not be a special point in the probability distribution, because the sum is over unrelated positive and negative terms of order unity. Hence, we may Taylor-expand the probability distribution around this point. For small $\sum C_i\ll 1$, it suffices to keep only the leading (constant) term~\cite{Wei87}. Thus, the prior probability for $\sum C_i\leq \epsilon\ll 1$ is of order $\epsilon$. 

We now integrate over these additional scanning parameters, and we require that $v$ remains below the catastrophic value for stability of nuclei beyond hydrogen. This modifies the prior distribution over $\tilde m$, yielding a marginalized distribution
\begin{equation}
\frac{dp}{d \log \tilde{m}} \; \propto  \left\{ \begin{array}{ll} \tilde m^q & (\tilde m \ll v) \\ 
\tilde{m}^{q-2} & (\tilde{m}  \gg v) \end{array} \right.
\label{eq-distabovev}
\end{equation}
that exhibits two different regimes. For $\tilde m \ll v$, the prior distribution is unmodified, with a probability force $q$ towards large $\tilde m$. But for $\tilde m \gg v$, the probability force is decreased to $q-2$, because of the statistical price that must be paid for fine-tuning the weak scale.

Hence we identify two behaviors:
\begin{enumerate}
\item If $q<2$, then the prior favoring large $\tilde{m}$ is too weak to overcome the statistical cost of fine-tuning (the accidental cancellation among the $C_i$ required for $\tilde{m} \gg v$). In this case, we would expect to discover natural supersymmetry ($\tilde m \sim v$, i.e., superpartners with mass of order the weak scale $v$). 
\item If $q>2$, then the multiverse force is sufficiently strong to favor runaway behavior for $\tilde{m}$, with superpartners many orders of magnitude above $v$, so that we expect to discover a finely-tuned theory of the weak scale.  
\end{enumerate} 
The first of these options is severely challenged by the failure, so far, of the LHC and other experiments to discover particles beyond the Standard Model. The second option would imply that we will never discover any, since $\tilde m \gg v$.

This analysis ignores the possible production of LSP dark matter. This may be appropriate. For example, the LSP might be unstable, or it might have a mass larger than the reheat temperature. Then the LSP would not contribute to the dark matter, and the above analysis would hold. One would then expect that the superpartners are out of reach of present experiments, and that the dark matter is associated with entirely different physics, such as an axion.

If LSP dark matter is produced, however, then our analysis so far is incomplete. The dark matter abundance will generally depend on $\tilde m$:
\begin{equation}
\zeta=\zeta(\tilde m)~.
\end{equation}
Thus, the probability distribution over $\tilde m$ will be modified by the baryon dilution factor of the causal patch measure of Eq.~(\ref{eq-mbxi}), yielding
\begin{equation}
\frac{dp}{d \log \tilde{m}} \; \propto  \left\{ \begin{array}{ll} \tilde m^q/[1+\zeta(\tilde m)] & (\tilde m \ll v) \\ 
\tilde{m}^{q-2}/[1+\zeta(\tilde m)]  & (\tilde{m}  \gg v) \end{array} \right.~~.
\label{eq-di}
\end{equation}
Below we show that baryon dilution creates a third regime---effectively, a catastrophe---whose threshold can set $\tilde m$. We will find that this scale is parametrically unrelated to the weak scale, but accidentally lies nearby.

\subsection{LSP Freeze-Out with Only $\tilde{m}$ Scanning }
\label{sec-LSP}

Here we specialize to a large class of supersymmetric theories where dark matter arises from freeze-out of the LSP.  We take the overall scale of superpartner masses, $\tilde{m}$, to scan, while keeping all other parameters fixed.  In particular the mass of superpartner $i$ is given by $\tilde{m}_i = A_i \, \tilde{m}$ with $A_i$ fixed.\footnote{If the $A_i$ are scanned, our analysis favors a hierarchy between $\tilde m$ and $m_X$, since this would allow the SUSY breaking scale to become large without incurring a penalty from the baryon dilution factor. The extent to which this hierarchy is realized depends on the prior distribution over the $A_i$, which is not known. Similarly, if the coupling strength $c=4\pi\alpha_{\rm eff}^2$ is scanned, it will be driven up to the unitarity bound~\cite{Griest:1989wd} unless the prior distribution disfavors this sufficiently.}  This class includes theories where the superpartners are at a single scale, with $A_i \approx 1$, and theories with a split spectrum.  For example, split supersymmetry \cite{ArkaniHamed:2004fb} has the fermionic superpartners much lighter than the scalar superpartners, so $A_{\rm ferm} \ll A_{\rm scal} \approx 1$.  With the discovery of a Higgs boson near 125 GeV, a spectrum based on anomaly mediation for gaugino masses \cite{Giudice:1998xp,Wells:2003tf} has become popular, variously called Spread \cite{Hall:2011jd, Hall:2012zp}, Pure Gravity Mediation \cite{Ibe:2011aa}, Mini-Split \cite{Arvanitaki:2012ps} and Simply Unnatural \cite{ArkaniHamed:2012gw}.  In these theories the gaugino masses have $A_a =  (b_a g_a / 16 \pi^2) \, ( M_*/\sqrt{3} M_{Pl})$, where $b_a$ and $g_a$ are the beta function coefficients and gauge couplings for gauge group $a$, and $M_*$ is a high mediation scale.  By taking $A_a$ to be fixed, the analysis of this sub-section also applies to these theories.

In the previous section, we showed that the mass of a freeze-out relic is proportional to the square root of the dark matter to baryon energy density ratio, by Eq.~(\ref{eq-zetaFO}): $m_X\propto\sqrt{\zeta}$. Our present assumptions link this mass to the SUSY breaking scale:
\begin{equation}
m_X = \tilde{m}_{LSP} = A_{LSP} \, \tilde{m}~.
\end{equation}
Hence,
\begin{equation}
\zeta \propto \tilde{m}^2~,
\end{equation}
and the probability distribution of Eq.~(\ref{eq-di}), which includes the baryon dilution factor of Eq.~(\ref{eq-mbxi}), takes the form
\begin{equation}
\frac{dp}{d \log \zeta} \; \propto  \;\; 
\frac{\zeta^{q/2-1}}{1 + \zeta} \hspace{0.3in} \mbox{for} \;\; \tilde{m}  \gg v~.
\end{equation}
This is identical to the distribution (\ref{eq-zetadist}) studied in Sec.~\ref{sec-whycomp}, with $n=q/2-1$.  Thus the Why Comparable problem is solved for most values of $q$ between 2 and 4. The distribution is then peaked near $\zeta =1$, as shown in Fig.~\ref{fig-xiprobs} and Fig.~\ref{fig-nrange}.

The mass of the LSP, $m_X=\tilde m_{LSP}$, is set by the freeze-out analysis of the previous section, and with our prediction $\zeta\sim O(1)$, is parametrically unrelated to the weak scale. By Eq.~(\ref{eq-FOscaling}),
\begin{equation}
\tilde m_{LSP}^2 \sim x_f^{-1} M_{\rm P} m_p \, c\, \eta .
\label{eq-FOscaling2}
\end{equation}
With order-one factors restored and the dependence on $\zeta\sim O(1)$ made explicit, this leads to  a statistical, environmental determination of the scale of supersymmetry breaking $\tilde{m} = \tilde{m}_{LSP} /A_{LSP}$:
\begin{equation}
\tilde m = \frac{0.5\, \mbox{TeV}}{A_{LSP}} \left( \frac{\alpha_{\rm eff}}{0.01} \right)  \sqrt{\zeta},~~\zeta\sim O(1).
\label{eq-mXFO2}
\end{equation}
If the spectrum is not highly split, $A_{LSP}\sim O(1)$, this predicts a Little Hierarchy with superpartners in the multi-TeV domain, somewhat above the weak scale.  In Split (or Spread) Supersymmetry, $A_{LSP}\ll 1$, only the fermionic superpartners (gauginos), including the dark matter particle, remain near the TeV domain, while the scalar superpartners are considerably heavier. Either way, there are superpartners with masses in the TeV domain that are parametrically unrelated to the weak scale and, in a variety of theories, these superpartners are accessible to the LHC and future colliders.

\subsection{Gravitino LSP Dark Matter}
\label{sec-gravitino}

The mass scale of the Standard Model superpartners is given by $\tilde{m} \sim F/M_{\rm mess}$ where $\sqrt{F}$ is the primordial scale of supersymmetry breaking and $M_{\rm mess}$ is the messenger scale.  All supersymmetric theories contain a gravitino of mass $F/M_{Pl}$; so unless $M_{\rm mess} \sim M_{Pl}$, the gravitino is expected to be the LSP.  (Special circumstances evade this conclusion~\cite{Hall:2011jd, Hall:2012zp, Ibe:2011aa, Arvanitaki:2012ps, ArkaniHamed:2012gw}.) 

Given the genericity of the gravitino as the LSP, it is important to investigate the implications for the observable superpartner mass scale $\tilde{m}$, if it is controlled not by the requirement of a natural weak scale as has long been assumed, but by $\zeta\sim O(1)$ as we have argued. 

Recently, it was shown that, under weak assumptions, if gravitinos are the dark matter $\tilde{m}$ cannot lie far from the TeV domain even if SUSY does not solve the hierarchy problem ~\cite{Hall:2013uga}. For simplicity we take all Standard Model superpartners to be comparable in mass so that the relevant parameter space is 3 dimensional $(T_r, \tilde{m}, m_{3/2})$.   To implement our solution of the Why Comparable problem we take $\tilde{m}$ to scan, and although $T_r$ and $m_{3/2}$ do not scan we allow them to take a wide range of values.   We take $m_{3/2} < \tilde{m}$ so the gravitino is the LSP, and $T_r > \tilde{m}$ so that the superpartners are cosmologically interesting.  We take $m_{3/2}$ sufficiently large, certainly greater than a keV, so that gravitinos could account for the dark matter.    With these restrictions $(T_r, \tilde{m}, m_{3/2})$ range over many orders of magnitude.

Although the gravitinos are sufficiently weakly interacting that they never reach thermal equilibrium, they are produced by three processes.   In Freeze-Out and Decay (FO\&D) the lightest Standard Model superpartner undergoes freeze-out, and at some later era decays to gravitinos; in Freeze-In (FI), when the temperature is of order $\tilde{m}$, Standard Model superpartners occasionally decay to give gravitinos; and finally gravitinos can be produced by scattering at high temperatures of order $T_r$ (UV).  For the usual Freeze-Out mechanism we found in (\ref{eq-FOscaling}) the parametric scaling behavior  $m_X^2 \sim ( M_{\rm P} m_p  \eta) \,  \zeta$.  For gravitino dark matter, if the production is dominated by (FO\&D, FI, UV) we find the scaling behaviors
\begin{equation}
\left( \tilde{m} m_{3/2}, \frac{\tilde{m}^3}{m_{3/2}}, \frac{\tilde{m}^2 T_r}{m_{3/2}} \right) \; \sim \; ( M_{\rm P} m_p  \eta) \,  \zeta.
\label{eq-gravDMscaling}
\end{equation}

Including numerical factors, these results can be used to predict the scale $\tilde{m}$ analogous to the freeze-out prediction of (\ref{eq-mXFO}).  When $r = m_{3/2} / \tilde{m}$ is not far below unity, FO\&D dominates giving
\begin{equation}
\tilde{m} = \frac{0.5}{\sqrt{r}} \left( \frac{\alpha_{\rm eff}}{0.01} \right)  \sqrt{\zeta} \; \mbox{TeV},
\label{eq-mFOD}
\end{equation}
with the TeV scale again emerging from $\sqrt{\eta \, M_{\rm P} m_p }$.  As $r$ drops so $\tilde{m}$ increases, but at $r=r_c$ the prediction for $\tilde{m}$ reaches a maximum as either FI or UV production dominates for $r < r_c$ giving
\begin{equation}
\tilde{m} = 25  \sqrt{ \frac{\alpha_{\rm eff}}{0.01} }  \sqrt{ \frac{r}{r_c} } 
 \left( \frac{\tilde{m}}{T_r} \right)^{1/4} \sqrt{\zeta} \; \mbox{TeV},
\label{eq-mFIUV}
\end{equation}
with $r_c = 5 \cdot 10^{-4} (\alpha_{\rm eff}/0.01)( T_r/\tilde{m})^{1/4}$.  For the FI case $T_r$ should be set equal to $\tilde{m}$.   Thus for $\alpha_{\rm eff} = 0.01$ the maximum value for $\tilde{m}$ is $25 (\tilde{m}/T_r)^{1/4}  \sqrt{\zeta} $ TeV, and this drops as $\sqrt{r}$ as $m_{3/2}$ is reduced further.

Thus scanning of the supersymmetry breaking scale, $\tilde{m}$, not only solves the Why Comparable problem but leads to a large class of theories, having a wide range of $T_r$ and $m_{3/2}$, with superpartners that may be accessible to LHC and future colliders.

\appendix
\section{Matter Mass in the Causal Patch}

Here we review how the matter mass $M$ in the causal patch at the time $t_{\rm obs}$ is computed~\cite{BouFre10d}. The Coleman-DeLuccia decay of an eternally inflating parent vacuum produces an open FRW universes~\cite{CDL} with metric $ds^2 = -dt^2 + a(t)^2 (d\chi^2 +\sinh^2\!\chi\, d \Omega_2^2)$. Here we consider the case where the daughter vacuum has positive cosmological constant, $\Lambda>0$; for an analysis of the $\Lambda<0$ case, see Ref.~\cite{BouFre10e}. We define
\begin{equation}
t_\Lambda =\sqrt{\frac{3}{\Lambda}}~,
\end{equation}
and we define $t_{\rm c}$ as the time when curvature begins to dominate over matter, or (if $t_\Lambda < t_{\rm c}$), when curvature {\em would\/} begin to dominate if $\Lambda$ was set to zero.

An approximate, continuous solution to the Friedmann equation is given by
\begin{equation} \label{eq:a}
a(t)\sim \left\{\begin{array}{ll}
t_{\rm c}^{1/3} t^{2/3}~,& t<t_{\rm c} \\
t~, & t_{\rm c}<t<t_\Lambda \\
t_\Lambda e^{t/t_\Lambda-1}~, & t_\Lambda < t~.
\end{array}\right.
\end{equation} 
For long-lived metastable de~Sitter vacua, the comoving radius of the causal patch can be computed to exponentially good accuracy by neglecting the decay and treating the vacuum as stable. Then the patch boundary is just the cosmological event horizon:
\begin{equation}
\chi_{\rm CP}(t) = \int_{t}^\infty \frac{dt'}{a(t')}~.
\end{equation}
Using the piecewise solution for the scale factor $a(t)$, one finds
\begin{equation}
\chi_{\rm CP}(t)\!\sim\!\! \left\{\begin{array}{ll}
\!\! 1+\log(t_\Lambda/t_{\rm c})+ 3[1-(\frac{t}{t_{\rm c}})^{1/3}], \  t<t_{\rm c} \\
\!\!1 +\log (t_\Lambda/t)~, \ \ \ \ \ \ \ \ \ \ \ \ \  \ \ \    t_{\rm c}<t<t_\Lambda \\
e^{-t/t_\Lambda}~, \ \ \ \ \ \  \ \ \ \ \ \ \ \ \ \ \ \ \ \ \ \ \ \ \ \ \ \ \ \     t_\Lambda<t.
\end{array}\right.
\label{eq-ccp}
\end{equation}
The above equations hold in the case where there exists a curvature-dominated era: $t_{\rm c}< t_\Lambda$. If $t_{\rm c}>t_\Lambda$, then curvature never dominates. (Our universe is in this parameter regime.) Results for the scale factor and comoving patch radius can be obtained from the above equations by setting $t_{\rm c} \to t_\Lambda$ and omitting the line corresponding to the curvature era ($t_{\rm c} <t< t_\Lambda$).

We are interested in the mass inside the causal patch at the time $t_{\rm obs}$, $M = \rho_{\rm m} a^3 V_{\rm com}= t_{\rm c} V_{\rm com} [\chi_{\rm CP}(t_{\rm obs})]$.  The comoving volume inside a sphere in hyperbolic space, $V_{\rm com}$, is $\chi^3$ for $\chi \lesssim 1$ (in the regime $ t_{\Lambda}<t_{\rm obs} $), and $e^{2\chi}$ for $\chi \gtrsim 1$ (i.e., for $ t_{\Lambda}>t_{\rm obs} $).  Combining expressions, one finds
\begin{equation}  
M\sim \left\{\begin{array}{ll}
\!\! 1/t_{\rm c}~, & t_{\rm obs}<t_{\rm c} < t_\Lambda \\
\!\! t_{\rm c} /t_{\rm obs}^2~,& t_{\rm c}<t_{\rm obs}<t_\Lambda \\
\!\! (t_{\rm c} /t_\Lambda ^{2}) e^{-3t_{\rm obs}/t_\Lambda}~, & t_{\rm c}< t_\Lambda< t_{\rm obs}\\
\!\! 1/t_{\Lambda}~, & t_{\rm obs}<t_{\Lambda} <t_{\rm c}   \\
\!\! t_{\Lambda}^{-1} e^{-3t_{\rm obs}/t_\Lambda}~, & t_\Lambda<t_{\rm obs}, \ t_\Lambda < t_{\rm c}.
\end{array}\right.
\label{eq-pcppos}
\end{equation}
For the purposes of this paper, the details of these expressions are irrelevant, which is why they are presented in an Appendix. What they show is that the total matter mass inside the patch does not depend on the dark matter to baryon density ratio $\zeta$.  Hence, the baryonic mass in the causal patch depends on $\zeta$ only through
\begin{equation}
M_b=\frac{1}{1+\zeta} M~.
\end{equation}

\acknowledgments We thank Roni Harnik and Yasunori Nomura for discussions. This work was supported by the Berkeley Center for Theoretical Physics, by the National Science Foundation (award numbers 1002399, 0855653 and 0756174), by fqxi grant RFP3-1004, by fqxi grant ``Multiverse Predictions for the LHC", by ``New Frontiers in Astronomy and Cosmology'', and by the U.S.\ Department of Energy under Contract DE-AC02-05CH11231.

\bibliographystyle{utcaps}
\bibliography{all}

\end{document}